# Taxon and trait recognition from digitized herbarium specimens using deep convolutional neural networks

**Authors**: Sohaib Younis [1,2], Claus Weiland [1], Robert Hoehndorf [6], Stefan Dressler [5], Thomas Hickler [1,4], Bernhard Seeger [2] & Marco Schmidt [1,3,5]

[1] Senckenberg Biodiversity and Climate Research Centre (SBiK-F), Frankfurt am Main, Germany

[2] Department of Mathematics and Computer Science, University of Marburg, Marburg, Germany

[3] Palmengarten der Stadt Frankfurt am Main, Frankfurt, Germany

[4] Department of Physical Geography, Goethe University, Frankfurt am Main, Germany

[5] Department of Botany and Molecular Evolution, Senckenberg Research Institute and Natural History Museum Frankfurt, Frankfurt am Main, Germany

[6] Computer, Electrical and Mathematical Sciences and Engineering Division, Computational Bioscience Research Center, King Abdullah University of Science and Technology, Thuwal, Saudi Arabia

e-mail for the corresponding author: Muhammad-Sohaib.Younis@senckenberg.de

## Abstract

Herbaria worldwide are housing a treasure of 100s of millions of herbarium specimens, which are increasingly being digitized in recent years and thereby made more easily accessible to the scientific community. At the same time, deep learning algorithms are rapidly improving pattern recognition from images and these techniques are more and more being applied to biological objects. We are using digital images of herbarium specimens in order to identify taxa and traits of these collection objects by applying convolutional neural networks (CNN). Images of the 1000 species most frequently documented by herbarium specimens on GBIF have been downloaded and combined with morphological trait data, preprocessed and divided into training and test datasets for species and trait recognition. Good performance in both domains is promising to use this approach in future tools supporting taxonomy and natural history collection management.

# Introduction

Herbaria have been the foundation for systematic botanical research for centuries, harbouring the type specimens defining plant taxa and documenting variability within them. Up to now, about 3000 herbaria have accumulated nearly 400 million specimens (Thiers 2017). These immense collections are rapidly becoming more accessible: the Natural History community presently experiences a major increase of digitization activities, with botany in Africa being among its pioneers with digitizing the type specimens in the context of the African Plants Initiative (Smith et al. 2011). While the availability of digitized plant specimens is constantly improving due to automated digitization streets and major collections are achieving almost complete coverage of their collections (Le Bras et al. 2017), simultaneously the opportunities for computer-based image recognition are also rapidly improving. Botanical applications include taxon recognition from photos in the context of citizen science, e.g. by Pl@ntnet (Joly et al. 2016) or iNaturalist (van Horn et al. 2017) or specially designed apps e.g. LeafSnap (Kumar et al. 2012). Up to now, there are only few and very recent applications on images of herbarium specimens (Unger et al. 2016, Carranza-Rojas et al. 2017, Schuettpelz et al. 2017). Our present approach includes taxon recognition as well as the recognition of morphological traits from herbarium specimens. Traits have so far only rarely been a subject in image recognition from plant images. There are however some approaches to extract leaf traits from plant images including specialized software and semiautomated workflows, that are comparatively work-intensive (e.g. Corney et al. 2012). We are focussing on taxa with a high number of images, in the context of trait recognition we focus further on African taxa, because of good availability of morphological trait data via a knowledge base generated within the framework of the Flora Phenotype Ontology (FLOPO; Hoehndorf et al. 2016) and African Plants - a photo guide (Dressler et al. 2014), but also because the existing taxon recognition presently works best for North American and European taxa and we are aiming to improve taxon recognition for a region, where taxonomic expertise and resources for identification are still less available and much needed.

Some deep learning approaches for recognising traits already exist that focus mainly on features like leaf count (Ubbens et al. 2017) and leaf tip/base (Pound et al. 2016) on selected model organisms to monitor performance of different cultivars or under different growth conditions. To our knowledge, this is the first study to deal with several traits in a large number of taxa, implying more abstraction in the concept of a trait and variability within a trait to be recognized.

# Materials and Methods

Our general approach in the recognition of taxa and traits from specimen images is to label the images with both of these informations, the taxon name being already included in the image data from the original data provider and the trait data being connected via the taxon name.

**Taxon data**: In order to resolve synonymies resulting from different concepts in our trait databases and the data accompanying the herbarium scans, we used the GBIF taxon backbone (GBIF 2017). Since scans have been found and downloaded via GBIF, names and taxon IDs from GBIF have already been attached to the images. Names from our trait data (FLOPO knowledge base / African Plants - a photo guide) have been matched to the GBIF backbone using the Global Names Resolver (http://resolver.globalnames.org/), only taxon name matches with a score > 0.9 have been used to label images with trait data.

**Trait data**: While herbarium scans in GBIF are labelled with a taxon name, traits are not directly connected with the scan and need to be linked via the taxon name, the approach is described above. This implies, that herbarium scans may or may not show all traits connected with the taxon. The trait dataset used in this study is from the multi-entry identification key of 'African Plants - a photo guide' (Brunken et al. 2008, Dressler et al. 2014) enhanced by trait data extracted from Floras using text-mining, annotated by a domain ontology (FLOPO) and combined in a knowledge base of plant traits which we consulted via a SPARQL endpoint (http://semantics.senckenberg.de/sparql).
In order to use traits, that are really shown on a herbarium scan, we focussed on a reduced set of leaf traits, considered to be recognizable in the majority of herbarium scans. These leaf traits include leaf arrangement, leaf structure, leaf form, leaf margin and leaf venation (Tab. 2)

**Images**: Our specimen imagery is from open access images contributed to the GBIF portal, a large part of these consisting of the MNHN vascular plant herbarium collection dataset in Paris (Le Bras et al. 2017). From several million digitized specimens on GBIF we downloaded scans of the 1000 species with most herbarium scans available via GBIF, consisting of a total of 830408 images. The distribution of scan images per species is heterogeneous, *Thymus pulegioides* L. with most images (5494) and *Orchis anthropophora* (L.) All. with least images (532).

For the trait recognition we extracted a subset of these (170 species / 152223 scans) with trait data available via the sources mentioned above. The dataset has been divided into 70% for training, 10% for validation and 20% for testing. The division is done uniformly for all species, regardless of the number of images belonging to it.

**Image Preprocessing**: All images in the dataset have been uniformly cropped and resized in portrait format after downloading them from GBIF. As shown in Figure 1, a typical herbarium sheet in this dataset contains a label with information on the collection event and the taxon as well as collector and number, in many cases further annotations by other scientists working with the collection object. It also contains bar codes on the top and bottom of the sheet and sometimes reference color bar on the sides. In order to reduce the background noise in the picture from the bar codes and color bars, they were uniformly cropped from the pictures, as suggested by Carranza-Rojas et al. (2017). One more reason to crop labels and notes is not to let the deep learning network learn from recognizing the characters in the notes but from the herbarium specimen. All images were cropped 7.5% from left and right in order to remove the reference color bars, 5% from top and 20% from bottom, to remove the bar codes and notes on the specimen. The images were then resized to 196 by 292 pixels, in order to preserve the aspect ratio of the sheet (Figure 1).

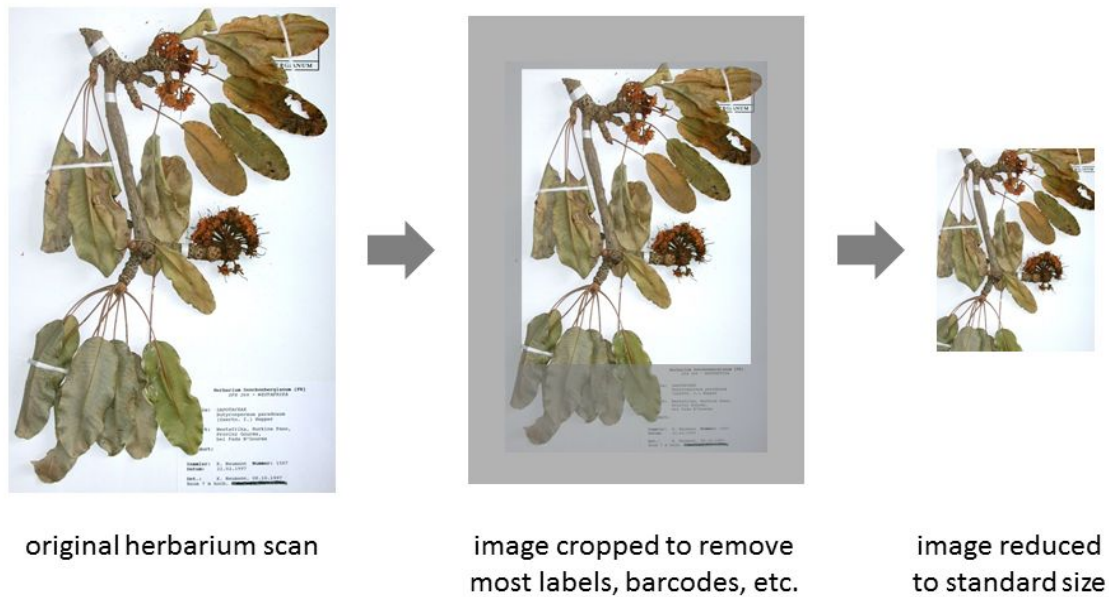

Figure 1. Image processing: herbarium scans as downloaded have been cropped and reduced to standard size in order to prepare them for treatment in deep learning algorithms.

**Deep Learning**: Deep learning is a part of machine learning methods for learning data representation. It processes the data in multiple layers, which leads to multiple levels of data abstractions, also known as features in image processing, for learning the representation. A typical deep learning network consists of multiple layers of artificial neural networks, inspired by biological neural networks. The main objective of a neural network is to learn a pattern or an approximation function based on its input. The connections have numeric weights that controls the signals between neurons, which can be tuned based on experience, making neural networks adaptive to input and capable of learning.

**Convolutional Neural Networks**: There are many types of deep learning networks but in our experiment we use convolutional neural networks (CNN) (LeCun 1995), as its main application is in image classification. The organization and connectivity of neurons in convolutional network is biologically inspired by animal visual cortex. Their success in the field of computer vision and image classification can be attributed to their need of relatively little or no preprocessing of images compared to other machine learning algorithms, which require the calculation of certain statistical properties of the image before learning or in some cases hand-engineered feature design. This is achieved by stacking multiple convolution layers, which apply a convolution operation to the input with a kernel, producing a feature map and passing it as input to next layer. As the network learns, the kernels in each layer are updated to improve the feature maps for the classification task. The initial layers in the network compute primitive features on the image such as corners and edges. The deeper layers use these features to compute more complex features consisting of curves and basic shapes and the deepest layers combine these shapes and curves to create recognizable shapes of objects in the image.

Convolutional neural networks have received attention recently due to their exceptional performance in ImageNet competitions. They have also demonstrated impressive results in recognition of plants in PlantCLEF challenges since 2015 (Goëau et al. 2017).

**Network Architecture:** The image recognition task was done by using a slightly modified ResNet model (He et al. 2016) implemented using the Tensorflow framework (Abadi et al. 2016). A typical residual convolution network uses a raw RGB image with dimensions of 225 by 225 as input with 3 three color channels (red, green and blue) and consists of blocks of convolution layers, a few pooling layers and a fully connected layer at the end, as shown in Table 1. It also contains skip connections to cater for vanishing gradients and degradation of information due to the depth of the network from large number of layers. These connections by pass the convolution layers in each block by carrying the output of the previous block and combining it with the output of the current block without any processing.

| Layer Type | Filter Size / Stride | Output Size |
|---|---|---|
| Convolution | 7 x 7 / 2 | 146 x 98 x 64 |
| Max Pool | 3 x 3 / 2 | 72 x 48 x 64 |
| Convolution | [1 x 1, 3 x 3, 1 x 1] x 3 | 72 x 48 x 256 |
| Average Pool | 2 x 2 / 2 | 36 x 24 x 256 |
| Convolution | [1 x 1, 3 x 3, 1 x 1] x 4 | 36 x 24 x 512 |
| Average Pool | 2 x 2 / 2 | 18 x 12 x 512 |
| Convolution | [1 x 1, 3 x 3, 1 x 1] x 6 | 18 x 12 x 1024 |
| Average Pool | 2 x 2 / 2 | 9 x 6 x 1024 |
| Convolution | [1 x 1, 3 x 3, 1 x 1] x 3 | 9 x 6 x 2048 |
| Average Pool | 9 x 6 / 9 x 6 | 1 x 1 x 2048 |
| Fully Connected (Softmax) | 1000 dense | 1000 |

Table 1. Sequence of layers in the ResNet and dimensions of features for each layer. Since the images for herbarium scans are resized to dimensions of 196 by 292, the last average pooling layer had to be modified to cater for the change in image size.

Figure 2 shows a block in ResNet, consisting of three convolutional layers. The first layer of 1x1 convolution reduces the number of filters for the next 3x3 layer, thus creating a bottleneck for the second layer. The third layer of 1x1 increases the number of filters again for the next block. The bottleneck in the block leads to a higher number of filters without increasing the model complexity. Each layer is followed by batch normalization and activation function, except the last layer where only batch normalization is used. The output of this batch normalization and previous block is added and then passed through an ELU activation function, which is then fed as input to the next block.

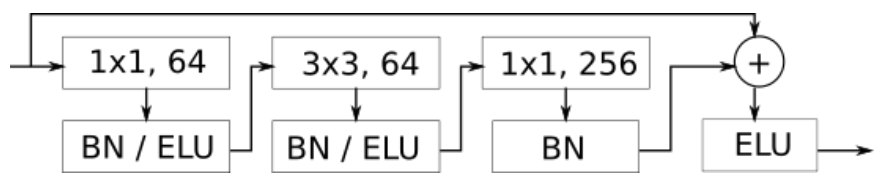

Figure 2. A bottleneck ResNet block

The model was implemented using the Tensorflow framework on a TITAN Xp GPU, we chose an approach using an Adam optimizer with Nestrov Momentum (Dozat 2016) in training. We started training the network with a batch size of 60 and learning rate of 1e-4 for 20k steps. It was then trained further with a batch size of 120 with learning rate of 5e-5 for 20k and then rate of 1e-5 for another 30k steps. The model processed the validation dataset every 500 steps during the last two stages of training to calculate the validation error. The training was early stopped if the validation error didn't decrease for the last 2500 steps.

For recognizing the traits, a smaller plain ResNet without bottleneck was used, since the total number of leaf traits in the dataset is limited to 19. As the scans have more than one unique trait, a sigmoid activation function was used in the last layer instead of softmax.

## Results

The network successfully predicted the correct species for the vast majority of images in the test dataset. Out of 165,689 test images, the network was able to correctly predict the species of 82.4% of the images correctly, in 96.3% of the images, the correct species was among the 5 most probable predictions. The most accurately predicted species was *Phlegmariurus phlegmaria* (L.) Holub, with 137 out of 138 images correctly predicted. The least accurately predicted species was *Rosa corymbifera* Borkh, with only 11 out of 157 images correctly predicted.

Generally probabilities for the correct species and accuracy values were high ( > 0.8) throughout the dataset, species with fewer images however more often had lower values (Figure 3).

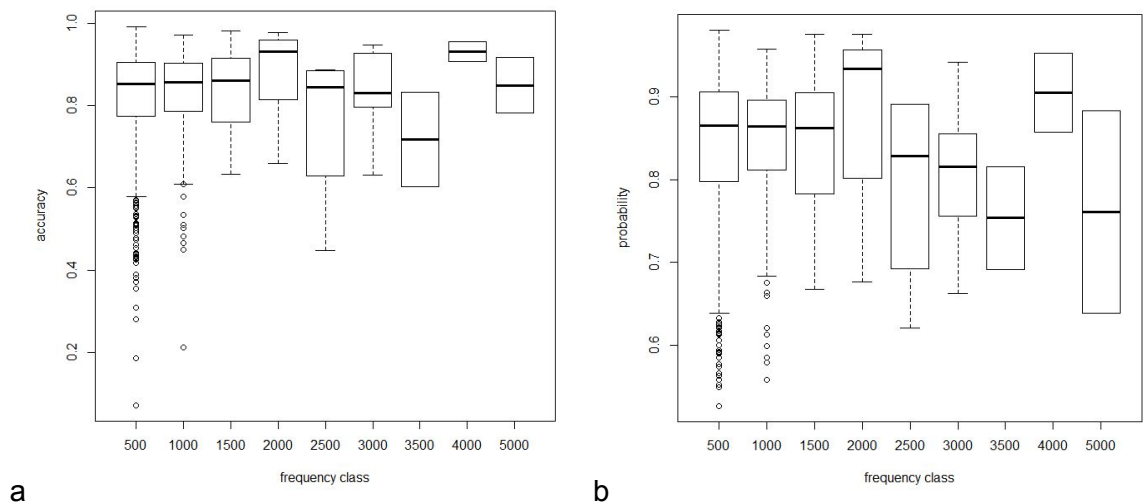

Figure 3. Species recognition from herbarium scans: performance measures depending on number of herbarium images (the frequency class includes all species with a number of images between x and x+500): a) boxplots of accuracy depending on number of test images per species . While the best-recognized species in each frequency class all have values >0.9, median values increase slightly, but especially minimum values increase strongly with number of images. b) boxplots of probability values for scans to be assigned to the correct taxon (mean probability of all test images of a taxon).

The network modified for trait recognition was able to predict, out of 30,374 test images, 89.6% of the traits successfully and including 30.9% of images with all the traits correctly. The most accurately predicted trait was Leaf - Structure (simple), with 21928 out of 22595 test images correctly predicted (Table 2). The least accurately predicted trait was Leaf - Structure (trifoliolate), with only 115 out 1266 test images correctly predicted.

| trait | state | OBO ID | # scans | accuracy(%) |
|---|---|---|---|---|
| Leaf - Arrangement | alternate | FLOPO:0001032 | 120514 | 92.98 |
| Leaf - Arrangement | opposite | FLOPO:0000420 | 34262 | 36.80 |
| Leaf - Arrangement | rosulate | FLOPO:0900066 | 37459 | 63.00 |
| Leaf - Arrangement | whorled | FLOPO:0002264 | 7550 | 44.61 |
| Leaf - Form | cordate | FLOPO:0900069 | 10378 | 29.81 |
| Leaf - Form | deeply lobed | FLOPO:0006834 | 28900 | 59.79 |
| Leaf - Form | oblong to linear | FLOPO:0000103 | 86644 | 81.00 |
| Leaf - Form | orbicular | FLOPO:0017811 | 8032 | 23.78 |
| Leaf - Form | ovate or elliptic etc. | FLOPO:0000286 | 91954 | 89.83 |
| Leaf - Margin | entire | FLOPO:0900073 | 118297 | 87.30 |
| Leaf - Margin | not entire (serrate or crenate etc.) | FLOPO:0900074 | 59148 | 72.50 |
| Leaf - Structure | palmately compound | FLOPO:0018499 | 2268 | 46.42 |
| Leaf - Structure | pinnately compound | FLOPO:0907004 | 46827 | 68.62 |
| Leaf - Structure | simple | FLOPO:0000693 | 128391 | 97.00 |
| Leaf - Structure | trifoliolate | FLOPO:0900067 | 8711 | 9.10 |
| Leaf Venation | palmate | FLOPO:0900070 | 17275 | 48.11 |
| Leaf Venation | parallel | FLOPO:0900072 | 40710 | 89.57 |
| Leaf Venation | pinnate | FLOPO:0000561 | 102663 | 90.35 |
| Leaf Venation | triplinerve | FLOPO:0900071 | 7372 | 21.0 |

Table 2. Leaf traits used in this study with OBO ID and number of herbarium scans labelled to have this trait. The last column shows the accuracy of predicted scans for each trait.

## Discussion

The only other similar approach for deep learning based species recognition of herbarium specimens with a large number of species is by Carranza-Rojas et al. (2017) with comparable results of 90.3% top 5 accuracy on 1204 species with 253,733 total images. The performance of species recognition from herbarium images in our study was very high, even close to 100% when the top 5 predictions were considered. However, this is largely due to high numbers of training data. Especially the number of badly recognized species was decreasing with an increase of digitized images available.

Trait recognition differed much between traits and seemed to depend to a large part on the number of training samples. However, comparatively poorly recognized traits such as trifoliolate leaves may also be difficult to recognize as a pattern, because they can be seen as a reduced form of both palmately and pinnately compound leaves and in their low number of leaflets have even some similarity to simple leaves.

Leaf venation patterns can generally not be recognized from images reduced to the size and resolution presently common in deep learning algorithms. Nevertheless, venation patterns have been successfully predicted in our study. This may well be due to other features closely connected with venation: parallel venation, e.g., often goes hand-in-hand with other characters of typical monocot leaves, like leaf shape and leaf margin. Triplinerve leaves on the other hand, are often quite similar in these aspects to palmately or pinnately veined leaves and have been predicted less successfully.

Leaf form was not as clearly defined in our trait dataset as other traits. Categories are relatively rough and not clearly separated from each other. Still, accuracy values were fairly high.

## Conclusion

The good performance of species and trait recognition from herbarium scans even for a dataset consisting of a large number of species is promising in the context of digitization activities worldwide. Automated species recognition may become a valuable tool for the taxonomist and technical staff facilitating identification and bringing possible misidentifications of herbarium specimens to the attention of the responsible curator. In the context of our study, identification performance is high enough to give valuable suggestions in the identification process and pre-identifications that may be used before assigning collection material to a taxon expert. We could well imagine the implication of deep learning algorithms in herbarium workflows, especially if material is digitized already at an early stage. However, we used species with a large number of herbarium images - for the majority of the c. 350,000 known species of plants, there are presently only very few or no images available. Further study is needed to include this "long tail of herbarium taxonomy". The application of deep learning on natural history collections is a very recent development and in the near future, further improvements may be expected especially from improved algorithms, higher availability of herbarium scans by ongoing digitization activities and faster computers allowing also larger image sizes to be processed and thereby to include more detail.

## Acknowledgements

We acknowledge the efforts of herbaria worldwide to digitize their collections and make scan images available via GBIF. We further want to thank all contributors to our trait data in the context of the FLOPO knowledge base and African Plants - a photo guide.

## Disclosure statement

## Funding details

SY, MS and SD received funding from the DFG Project *Mobilization of trait data from digital image files by deep learning approaches* (grant 316452578). Parts of RH's & CW's work were funded by the National Bioscience Database Center (NBDC) and the Database Center for Life Science (DBCLS) Biohackathon 2017 grants. We gratefully acknowledge the support of NVIDIA Corporation with the donation of the TITAN Xp GPU to CW used for this research.

## Notes on Contributors

**Sohaib Younis** is a computer scientist at Senckenberg Biodiversity and Climate Research Center with focus on deep learning and image processing.
Contributions: convolutional network modeling, image preprocessing, description of results and preparation of manuscript

**Claus Weiland** is scientific programmer at SBIK-F's Data & Modelling Centre with main interests in large-scale machine learning, trait semantics and scientific data management
Contributions: Flora Phenotype Ontology and knowledge base, design of the GPU platform, data analysis and preparation of the manuscript.

**Robert Hoehndorf** is assistant professor for computer science at the King Abdullah University of Science and Technology. His research interests are artificial intelligence, knowledge representation, biomedical informatics, ontology.
Contributions: stimulation and concept of study, design and implementation of workflow for the Flora Phenotype Ontology and knowledge base.

**Stefan Dressler** is curator of the phanerogam collection of the Herbarium Senckenbergianum Frankfurt/M., which includes its digitization and curation of associated databases. Taxonomically he is working on Marcgraviaceae, Theaceae, Pentaphylacaceae and several Phyllanthaceous genera.
Contribution: Trait data

**Thomas Hickler** is head of SBIK-F's Data & Modelling Centre and Professor for Biogeography at the Goethe University Frankfurt. He is particularly interested in interactions between climate and the terrestrial biosphere, including potential impacts of climate change on species, ecosystems and associated ecosystem services.
Contribution: Preparation of manuscript, comprehensive concept of study within biodiversity sciences.

**Bernhard Seeger** is professor of computer science systems at the Philipps University of Marburg. His research fields include high performance database systems, parallel computation and real time processing of high-throughput data with a focus on spatial biodiversity data.
Contribution: Provision of support in machine learning and data processing.

**Marco Schmidt** is a botanist at Senckenberg Biodiversity and Climate Research Center (SBIK-F) with a focus on African savannas and biodiversity informatics (e.g. online databases like *African Plants - a photo guide* and *West African vegetation)* and is working at Palmengarten's scientific service, curating living collections and collection databases. Contributions: concept of study, workflow, taxon and trait data, preparation of manuscript